\documentclass[aps,pra,twocolumn,groupedaddress,doublespace,superscriptaddress]{revtex4-1}
\usepackage{epsfig,subfigure,subfloat,graphicx,amssymb,latexsym,amsthm
,amsmath,natbib,slashed,dsfont,color,nicefrac}
\usepackage{color}

\usepackage{hyperref}
\DeclareRobustCommand{\baselinestretch{2.2}}

\hypersetup{
    colorlinks,
    citecolor=blue,
    filecolor=black,
    linkcolor=red,
    urlcolor=blue
}
\hypersetup{linktocpage}

\newcommand{\beq}{\begin{equation}}
\newcommand{\eeq}{\end{equation}}
\newcommand{\beqa}{\begin{eqnarray}}
\newcommand{\eeqa}{\end{eqnarray}}

\newcommand{\ben}{\begin{enumerate}}
\newcommand{\een}{\end{enumerate}}
\newcommand{\bit}{\begin{itemize}}
\newcommand{\eit}{\end{itemize}}
\newcommand{\bpm}{\begin{pmatrix}}
\newcommand{\epm}{\end{pmatrix}}

\newcommand{\pad}{\partial}

\renewcommand{\baselinestretch}{1.2}

\def\V{{\cal V}}
\def\PP{{\cal P}}
\def\N{{\cal N}}
\def\H{{\cal H}}

\def\Ord{{\cal O}}

\def\lt{\left}
\def\rt{\right}

\def\bra#1{{\langle #1|}}
\def\ket#1{{| #1\rangle}}

\def\bracket#1#2 {\mathinner{\langle{#1}|{#2}\rangle}}

\def \im {{\cal I}m}
\def \re {{\cal R}e}

\def \vphi {\varphi}

\def \dl {\delta}

\def \gm {\gamma}

\def \sg {\sigma}

\def \nb {\nabla}

\def\->{\rightarrow}

\def\br{{\bs r}}
\def\bk{{\bs k}}

\def\bp{{\bs p}}

\def\wt{\widetilde}

\def \bs {\boldsymbol}

\def \. {\cdot}

\def \. {\cdot}

\global\long\def\un{{\mathds1}}

\begin{document}


\title{Nonlinear wave dynamics in honeycomb lattices}


\author{Omri Bahat-Treidel}
\affiliation{Department of Physics, Technion-Israel Institute of
Technology, Technion City, Haifa 32000, Israel}
\author{Mordechai Segev}
\affiliation{Department of Physics, Technion-Israel Institute of
Technology, Technion City, Haifa 32000, Israel}


\today

\begin{abstract}  

We study the nonlinear dynamics of wave packets in honeycomb lattices, and show that, in quasi-1D configurations, the waves propagating in the lattice can be separated into left-moving and right-moving waves, and any wave packet composed of left (or right) movers only does not change its intensity structure in spite of the nonlinear evolution of its phase. We show that the propagation of a general wave packet can be described, within a good approximation, as a superposition of left and right moving self-similar (nonlinear) solutions.  Finally, we find that Klein tunneling is not suppressed due to nonlinearity.
\end{abstract}

\pacs{}

\maketitle



The vast interest in honeycomb lattices, which  started more than 25 years ago in condensed matter by showing that electron waves obey the massless Dirac equation  \cite{PhysRevLett.53.2449,PhysRevLett.53.742.2}, has recently spread to numerous other fields. Examples range from electromagnetic waves in photonic crystals  \cite{PhysRevLett.100.013904, PhysRevA.78.033834, PhysRevA.75.063813,PhysRevB.78.045122,PhysRevB.80.155103} and waveguide arrays \cite{PhysRevLett.98.103901, Bahat-Treidel:08, PhysRevLett.104.063901}, to cold atom in optical lattices  \cite{PhysRevLett.99.070401, zhu-2007-98, PhysRevB.77.235107, 
PhysRevA.80.043411} and more. However, despite having the same honeycomb-structured  potential, there are also some very important differences between these various systems, mainly because the interactions between the waves are different in nature. Namely, in graphene, the electrons have coulomb interaction and spin exchange, whereas EM waves can interact via nonlinearity of different types (Kerr, saturable, etc. ), and cold atoms can be either bosons or fermions and display dipolar or nonlocal interactions. Naturally, it  would be very interesting to study the effects of the different types of interactions on the phenomena  associated with the linear (non-interacting) regime. For example, it was found that Klein tunneling in honeycomb lattices  \cite{Katsnelson_2006} is strongly suppressed by coulomb interaction  \cite{PhysRevB.80.235423}. Would interactions in other nonlinear systems also suppress Klein tunneling or perhaps some types of interactions preserve this extraordinary phenomenon ?

Here, we study the dynamics of waves in honeycomb lattices, in the presence of
Kerr nonlinearity, which applies to photonic crystals, waveguide arrays and
Bose-Einstein condensates (BEC). We focus on quasi-1D wavepackets: wave
packets that are very wide in one direction and quite narrow in the other transverse
direction. We find self-similar closed-form solutions for the nonlinear Dirac
equation, i.e., solutions whose intensity structure remains unchanged during
the propagation, except for a shift of the center. The spatial form of these
solution can be completely arbitrary, as long as they are either left movers or right
movers only. Moreover, we show that the propagation of a general wave packet in
the honeycomb lattice can be described using superposition principle, to within
a very good approximation, even in the presence of significant nonlinearity.
Finally, we reexamine Klein tunneling in the presence of nonlinearity, and
find that, as opposed to the electronic case, Klein tunneling is unaffected.

\begin{figure}[tb]
    \center
    {\includegraphics[width=0.225\textwidth]{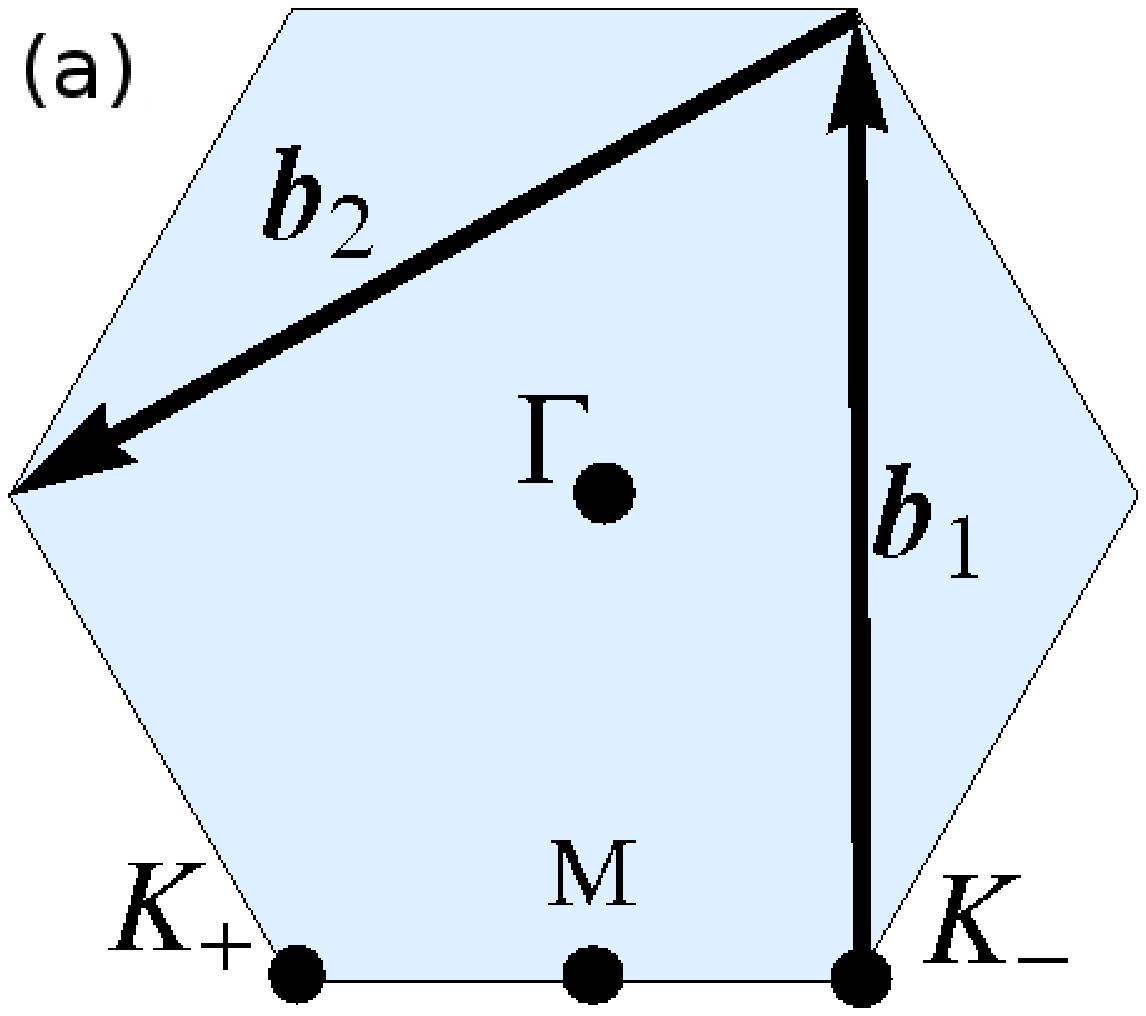}
    \includegraphics[width=0.215\textwidth]{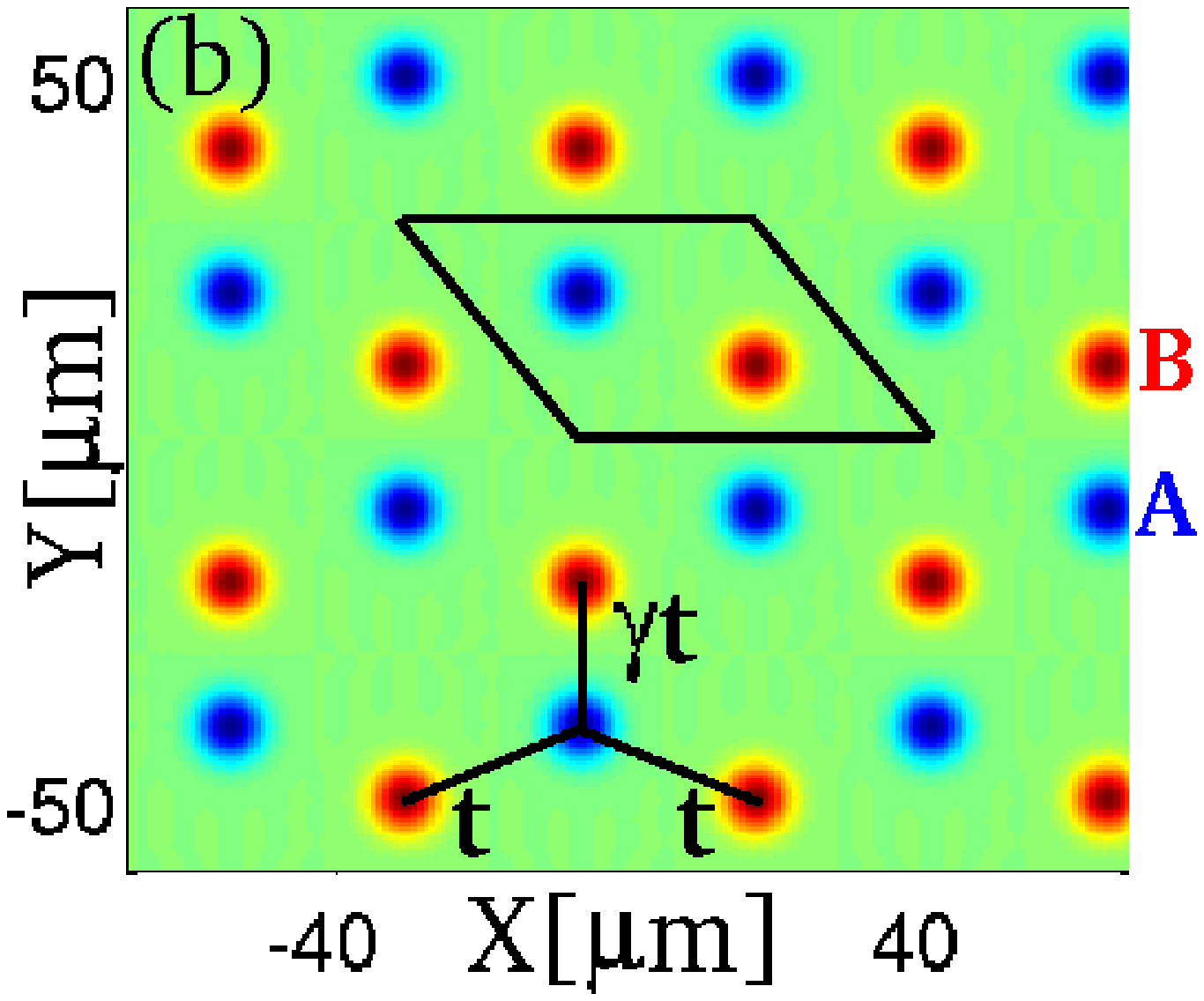}  
    }%
    \caption {(a) The first Brillouin zone with the high symmetry points.
    (b) honeycomb lattice that has two sites in a unit cell.}
    \label{lattice and BZ}
\end{figure}

For concreteness, we analyze here a honeycomb photonic lattice displaying the Kerr nonlinearity (most common optical nonlinearity). The paraxial propagation of a monochromatic field envelope  $\Psi$ inside a photonic lattice exhibiting the Kerr non-linearity is described by
\beq
    i \pad_z \wt \Psi  = -\frac{1}{2k_m}\nb^2_{\perp} \wt \Psi -
    \frac{k_m\dl n(x,y)}{n_0} \wt \Psi - \frac{k_m}{n_0}n_2 |\wt \Psi|^2 \wt \Psi,
    \label{nls}
\eeq %
where  $\dl n(x,y)$ is the modulation in the refractive index defining the lattice (Fig. \ref{lattice and BZ} (a)),  $k_m$ is the wave-number in the medium,  $n_0$ the background refractive index, and  $n_2$ is the Kerr coefficient. 
The sign of  $n_2$ determines the type of nonlinearity, where  $n_2 > 0$ 
corresponds to a focusing nonlinearity (attractive interactions in the context of BEC). 
The term  $k_m \dl n/n_0$ is referred to as the optical potential. It is convenient to transform the above equation to dimensionless form
\beq
	i \pad_z \Psi = -\nb_{\perp}^2 \Psi - \V(\br) \Psi - U |\Psi |^2 \Psi,
\eeq
where the coordinates are measured in units of  $k_m^{-1}$.

Since the Kerr nonlinearity is {\em local}, each pseudo-spin (sub-lattice) component is expected to be affected only by its own intensity.
We can now write the field as a two-component field,  $\Psi^{\dag} \equiv ( \psi_A ~~ \psi_B ) $, where  $\psi_A, \psi_B$ are the amplitudes of the electric field on the two sub-lattices. By projecting on the Wannier states of the two lowest bands, it is possible to describe excitations close to the Dirac points by \cite{PhysRevA.79.053830, nonlinear_dirac}.
\beq
	i \pad_z \Psi = \H_0  \Psi - U \hat n \Psi, \quad \hat n \equiv \text{diag} (|\psi_A |^2,|\psi_B|^2 ),
	 \label{general nonlinear honeycomb eq}
\eeq
where  $\H_0$ describes the linear dynamics in the system, which can be elegantly (and rather accurately) expressed using the tight-binding approximation 
\beq
	\H_0 = \re\{\vphi(\bk)\}\otimes \sg_x + \im\{\vphi(\bk)\}\otimes \sg_y - V_{\text{ex}}(\br)\otimes \un,
\eeq
where $\vphi(\bk) = \sum_{j=1}^3 t_j \exp(i\bs{\dl_j}\cdot \bk)$, $t_j$'s are the hoping parameters, $\bs{\dl}_j$ are the vectors connecting the nearest neighbors, and  $V_{\text{ex}}(\br)$ is some additional external potential \cite{PhysRevLett.104.063901,dietl:236405,PhysRevB.80.153412}. Considering uniaxial deformations  $t_1 = t_2 = t,~t_3 = \gm t$, and expanding  $\vphi$ around one of the Dirac points (say  $\bs {K}_+ $ )  $\vphi(\bk = \bs{K}_+ + \bp)\simeq v_x p_x - i v_y p_y $, we find that, to leading order, Eq.({\ref{general nonlinear honeycomb eq}}) reduces to the nonlinear Dirac equation
\beq
	i \pad_z \Psi = -\lt[ i\sum_{j=x,y} v_j \pad_j \otimes \sg_j 
	+ V_{\text{ex}}(\br)\otimes \un \rt]\Psi - U \hat n \Psi,
\eeq
where $v_x = \sqrt{1-(\gm/2)^2},~v_y = \sqrt{3}\gm/2$, and $\bp$  is measured in units of  $a^{-1}$ where  $a$ is the lattice constant.

In what follows we consider a quasi-1D scenario, meaning that we consider wave packets that are very broad in one direction and narrow in the other. To leading order in the momentum, the Dirac Hamiltonian is isotropic and all the directions (with respect to the lattice) are equivalent (except for a numerical factor  $v_j$). Therefore, we consider a wave packet moving in the x-direction and hence put  $p_y=0$. The resulting propagation equation is
\beq
	i \pad_z \Psi  = -\lt[iv_x \pad_x \otimes \sg_x + V_{\text{ex}}(x)\otimes \un \rt]\Psi - U \hat n \Psi.
	\label{1D dirac eq}
\eeq
The solutions of the linear equation are eigenstates of $\sg_x$. Since the two components of the eigenvectors of  $\sg_x$ differ only in their phases  $\bra{\pm} = \bpm 1 & \pm 1 \epm$, the nonlinear term is proportional to the identity operator (in the pseudo-spin space). Therefore, the spinors solving the linear equation also solve the nonlinear equation. It is therefore sensible to look for solutions of the form 
\beq
	\Phi^{T}_{\text{trial}} = \bpm f(x,z) ~ & \pm f(x,z) \epm. \label{trial solution}
\eeq
Substituting (\ref{trial solution}) into (\ref{1D dirac eq}) (setting  $V_{\text{ex}}(x)=0$) we obtain
\beq
	 \pad_z f(x,z) = \mp v_x \pad_x f(x,z) + iU \lt|f(x,z) \rt|^2 f(x,z).
\eeq
This equation has general solutions of the form
\beq
	f(x,z) = g(x\pm v_x z)\exp \lt[i U | g(x\pm v_x z)|^2 z \rt], \label{exact nonlinear solution}
\eeq
Where   $g(\xi)$ is some {\em arbitrary} function. The 'up' state,  $\ket{+}$, corresponds to a left moving solution $g(x+v_x z)$, whereas the 'down' state,  $\ket{-}$, corresponds to  a right moving solution $g(x-v_g z)$. It is useful to think of the solutions,  $g(x\pm v_x z)$, as linear combinations of plane waves. The right (left) moving solution is a combination of waves the move to the right (left) and satisfy  $\beta_z = -v_x p_x$ ($\beta_z = v_x p_x$). 

This form of solutions has profound implications:
\ben
	\item The intensity of {\em any} wave packet is unchanged 
	throughout propagation, except for some drift in its absolute position.
	\item The wave packet does not experience any broadening or 
	narrowing as a consequence of the nonlinearity, as generally happens in 
	other nonlinear systems.
	\item The sign of the nonlinearity is irrelevant for the intensity structure. 
	It affects only the phase of the wave packet.
	\item Since $g(\xi)$ is arbitrary any additional noise is simply some other function $\wt g(\xi)$ which is also 
	a self-similar solution. Hence the phenomenon of modulation instability is impossible in this system. That
	is, all wave packets are inherently stable.
\een

\begin{figure}[tb]
    \center
    {\includegraphics[width=0.3\textwidth]{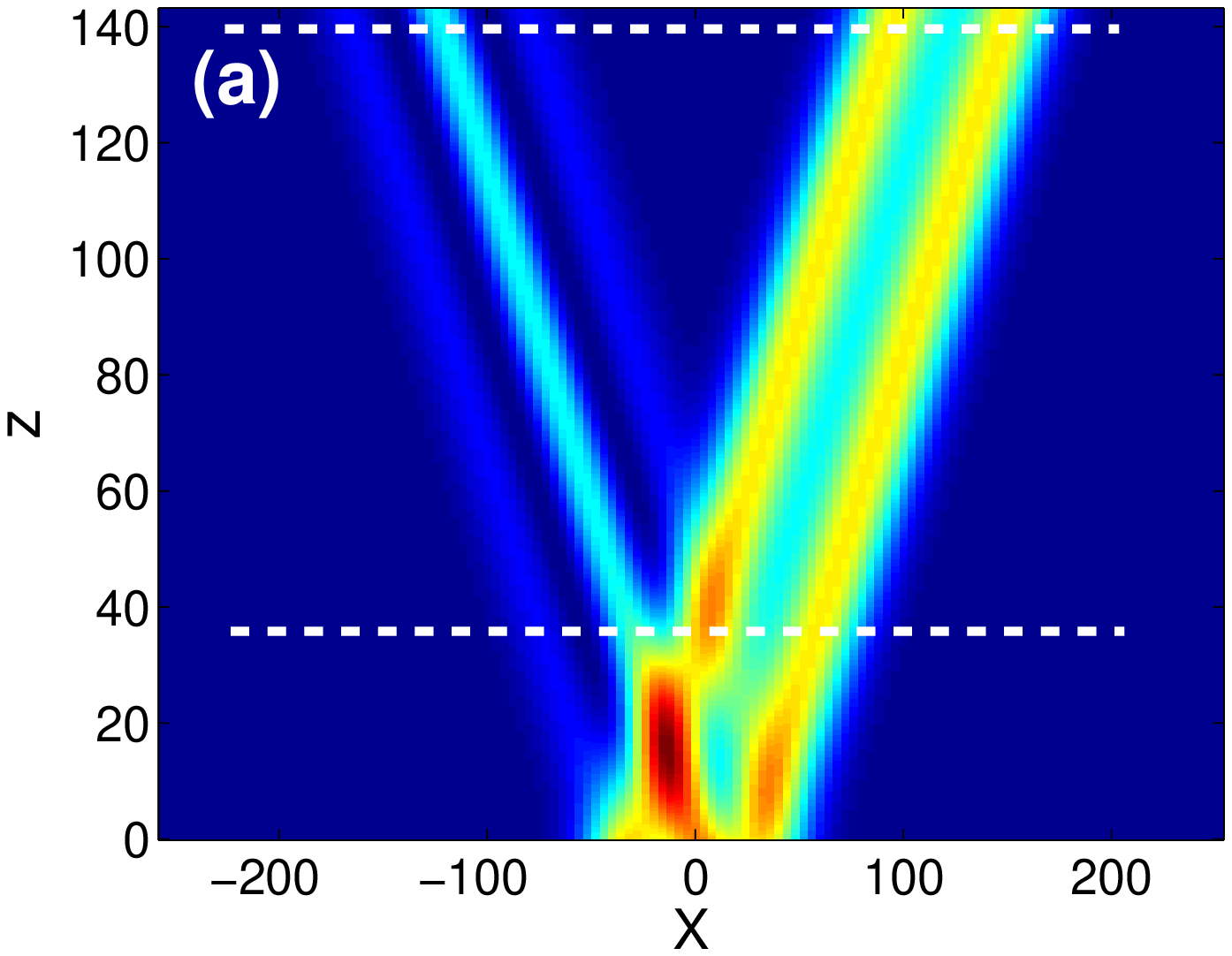}
    \includegraphics[width=0.3\textwidth]{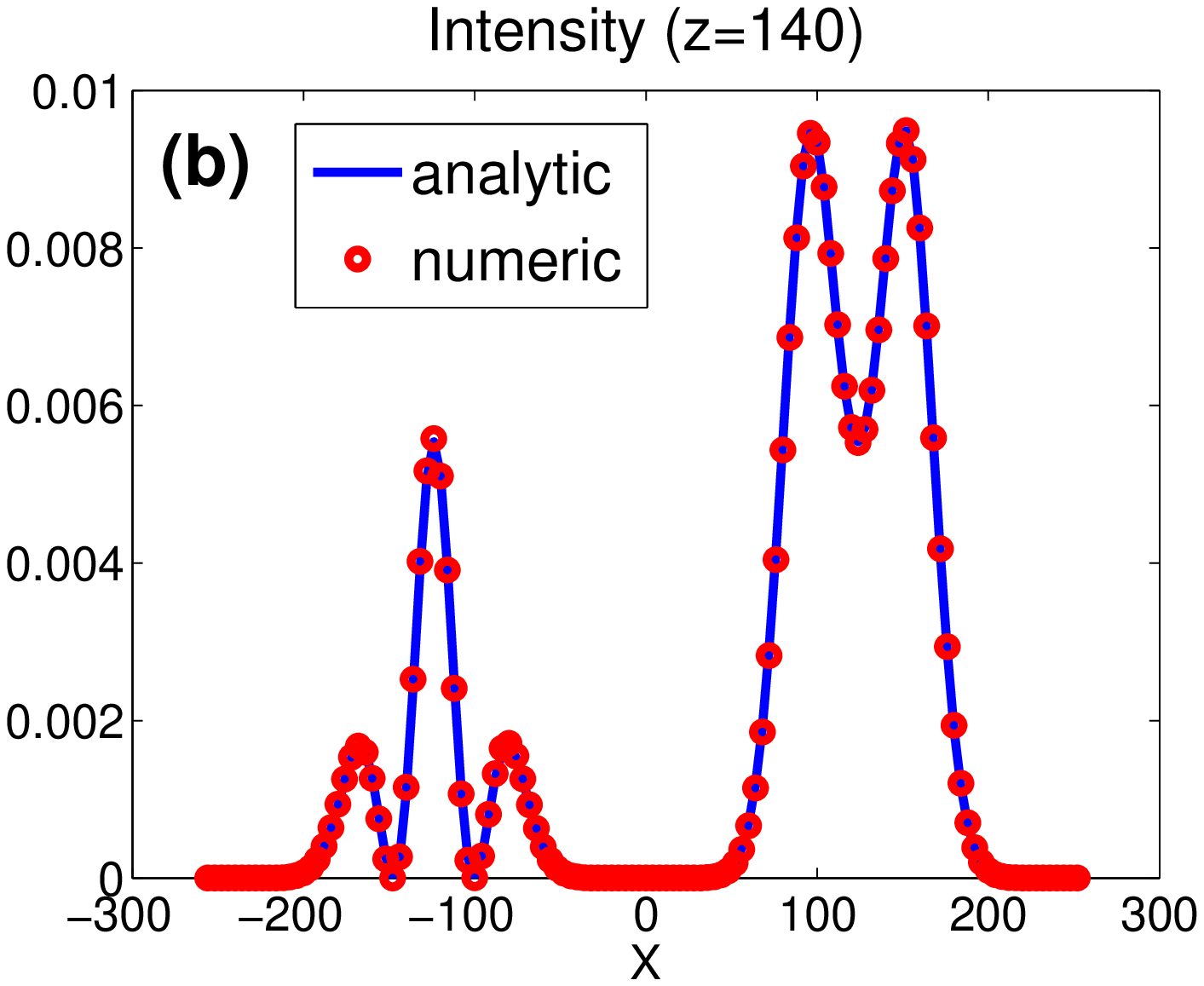}
    \includegraphics[width=0.3\textwidth]{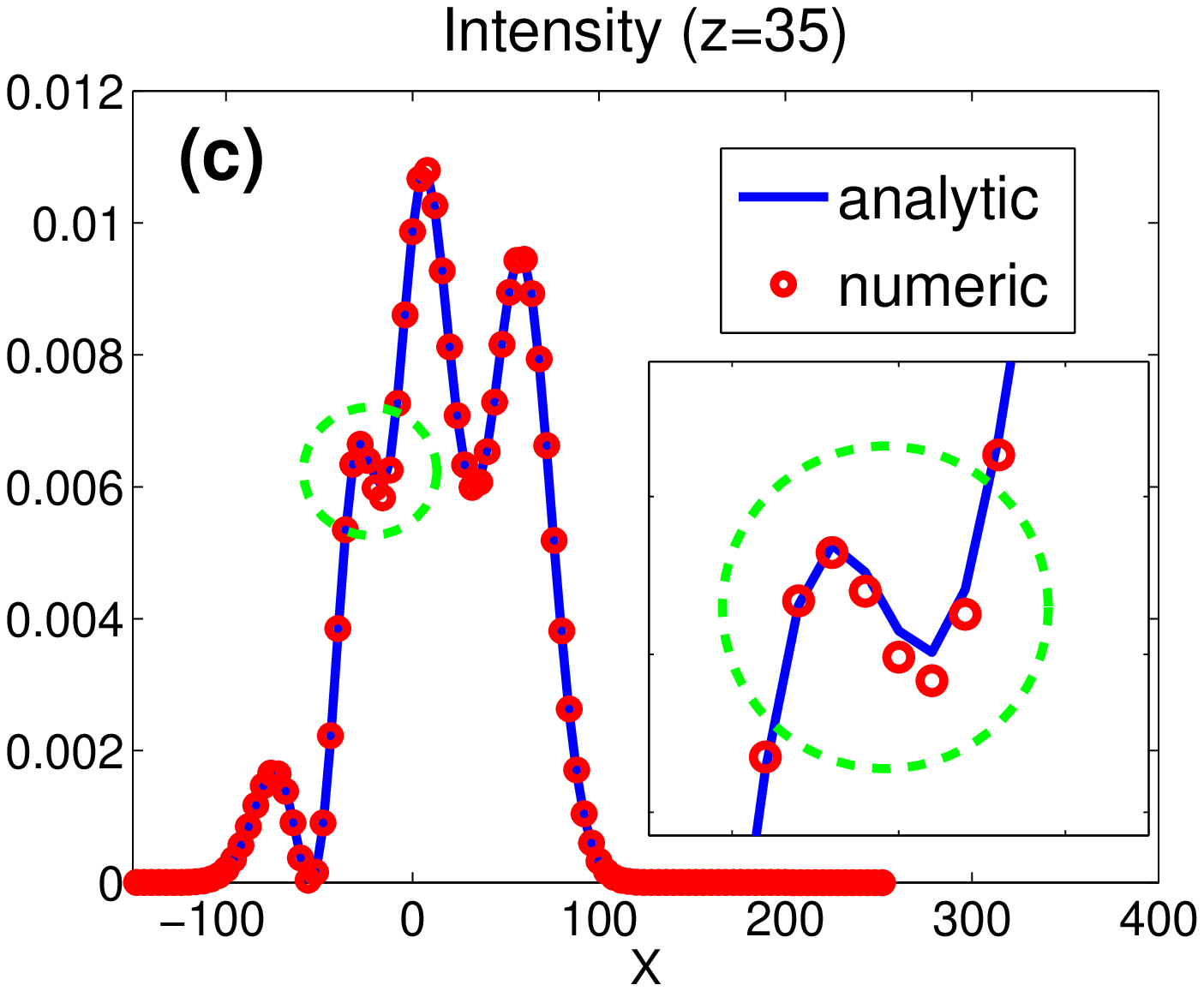}
    }%
    \caption {Simulated propagation of general wave packet (a), and comparison between the simulation 
    and the superposition of the right and left moving solutions at $z=140$ (b) and at $z=35$ (c). Inset: the greatest discrepancy 
    between the numeric and the analytic solutions.}
    \label{superposition}
\end{figure}

However, the most general wave packet is not composed of right (left) movers only. Rather, it is made of a superposition of the two, 
\beqa
	\Psi(x,0) = \bpm f_A(x) \cr f_B(x)\epm = h_+(x,0)\ket{+} + h_-(x,0)\ket{-},~~
\eeqa
where   $h_{\pm}(x,0) \equiv \lt(f_A(x,0) \pm f_B(x,0) \rt)/2$. It would be very unusual if the dynamics of a general wave packet propagating in the nonlinear honeycomb lattice would be the sum of the right-moving and left-moving self-similar solutions:
\beq
	\Psi_{\text{sup}}(x,z) = h_+(x,z)\ket{+} +  h_-(x,z)\ket{-}
\eeq
where  $h_{\pm}(x,z)$ are of the form  (\ref{exact nonlinear solution}).
And indeed, even though the equation is nonlinear - where in general the sum of two solutions is not a solution, in this case such superposition turns out to be an excellent approximation. The reason is that the terms that 'spoil' the superposition are products of  $h_+$ and  $h_-$ (and their c.c). Since any physical wave packet has a finite width,  $\dl x$, after large enough propagation distance - the overlap between the $h_+(x-v_x z)$ and  $h_-(x+v_x z)$ is negligible (since they move in opposite directions). Therefore, for   $z \gg \dl x$ a general wave packet evolves to a superposition of the left and right moving nonlinear solutions given in  (\ref{exact nonlinear solution}). We demonstrate the validity of this superposition principle for the nonlinear dynamics in honeycomb lattices by solving Eq.(\ref{1D dirac eq}) numerically  with initial condition  $ f(x,0) = \N_f\exp(-x^2/\sg^2),~g(x,0) =\N_g x^2 \exp(-x^2/\sg^2)$, where  $\N_f,\N_g$ are normalization constants, set $U=0.5$ and  $v_x = \sqrt{3}/2$, and compare to the superposition of the analytic solutions. We find that the agreement is excellent at large propagation distance ( $z\gg\dl x$) where the two solutions are spatially separated (Fig. 2 \ref{superposition} (b)), while even at short distances there is only very small discrepancy between the two  ( $z \lesssim \sg$) (Fig. \ref{superposition} (c) ). 
%

%

Next, we examine this unique wave dynamics when taking into account the fact that  $\beta_z$ is not really linear in  $p_x$. We do that by including higher order terms in the expansion of  $\vphi(\bk)$. We emphasize that  $\vphi(\bk)$ is not isotropic (even for a non deformed lattice, i.e.,  $\gm=1$) beyond the leading terms, and therefore one has to specify the direction with respect to the lattice. This anisotropy can be used to distinguish between excitations residing  in the different Dirac cones, which suggests applications in   controlling an electronic devices  \cite{Valley_filter_Beenakker,PhysRevLett.100.236801,0953-8984-21-4-045301}. We proceed to study quasi 1D dynamics in the x-direction, considering the next term the kinetic terms is
\beq
	\vphi(\bp) = v_x p_x + \frac{\gm}{8}p_x^2,
\eeq 
where the significance of the quadratic term can be tuned by controlling the deformation of the lattice. We emphasize that, as we approach  $\gm \to 2$, a qualitative difference arises between the different directions:  $v_x \to 0$ whereas  $v_y$ is finite and nonzero, as was demonstrated for Klein tunneling \cite{PhysRevLett.104.063901}. We find that, when the quadratic term is significant enough such that in the absence of nonlinearity ( $U=0$) the wave packet experiences noticeable diffraction, the nonlinearity has its usual effects, that is,   $U>0$ cause focusing of the beam, whereas  $U<0$ enhances the beam broadening. However, we emphasize that in honeycomb lattices there is a significant range of parameters where the quadratic terms are negligible and the exotic wave dynamics can be observed in experiments. 

\begin{figure}[]
    \center
    {\includegraphics[width=0.23\textwidth]{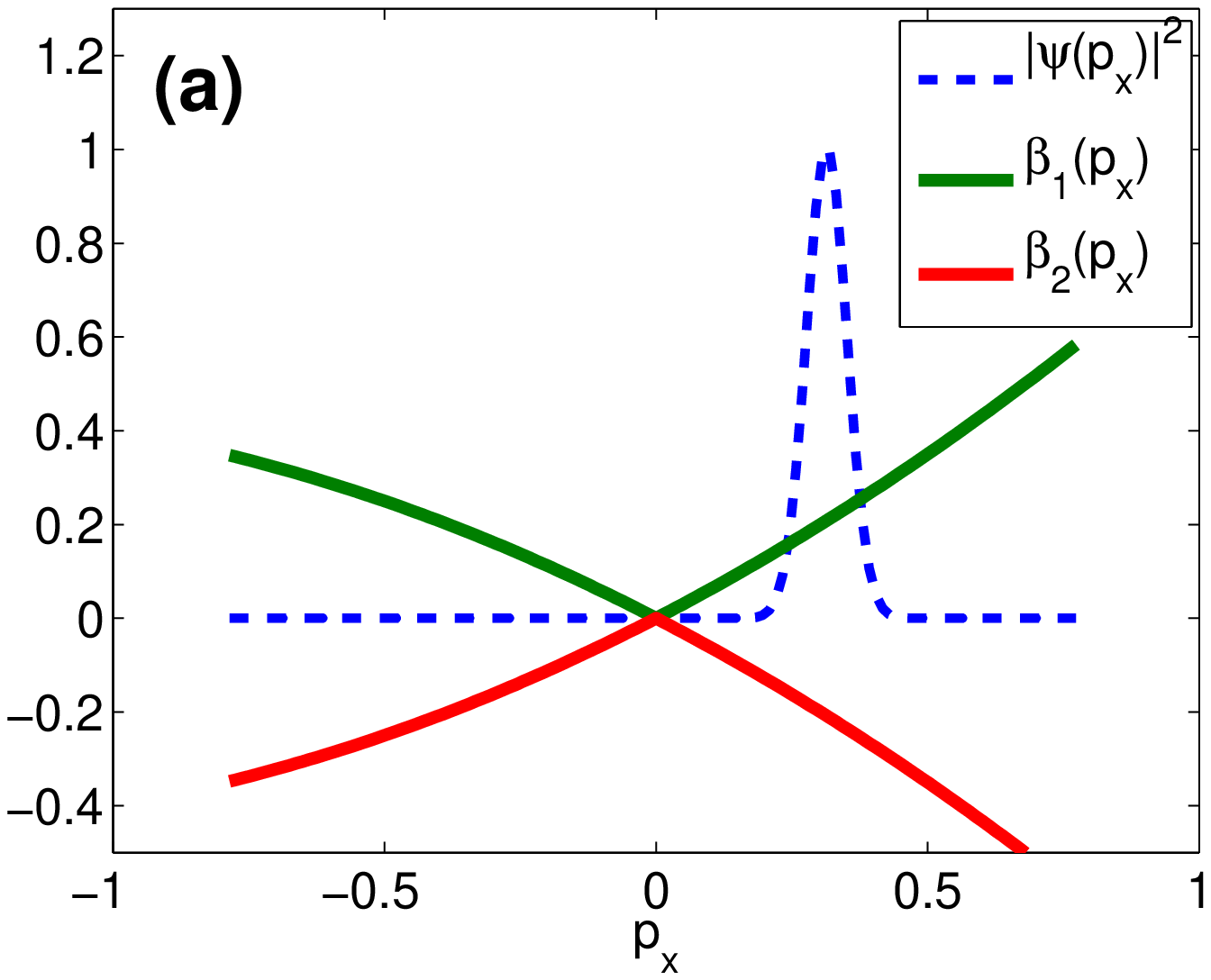}
    \includegraphics[width=0.23\textwidth]{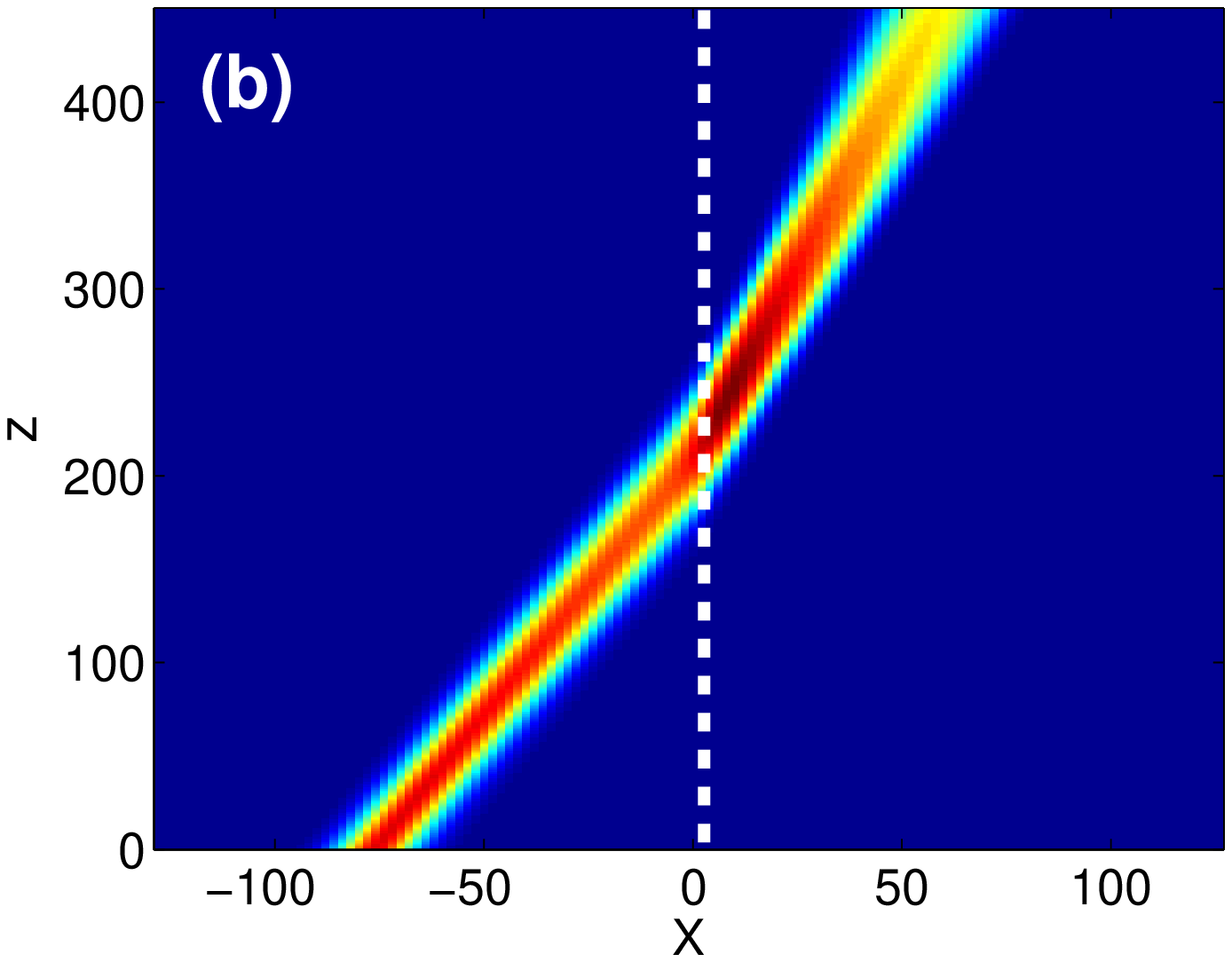}
    \includegraphics[width=0.23\textwidth]{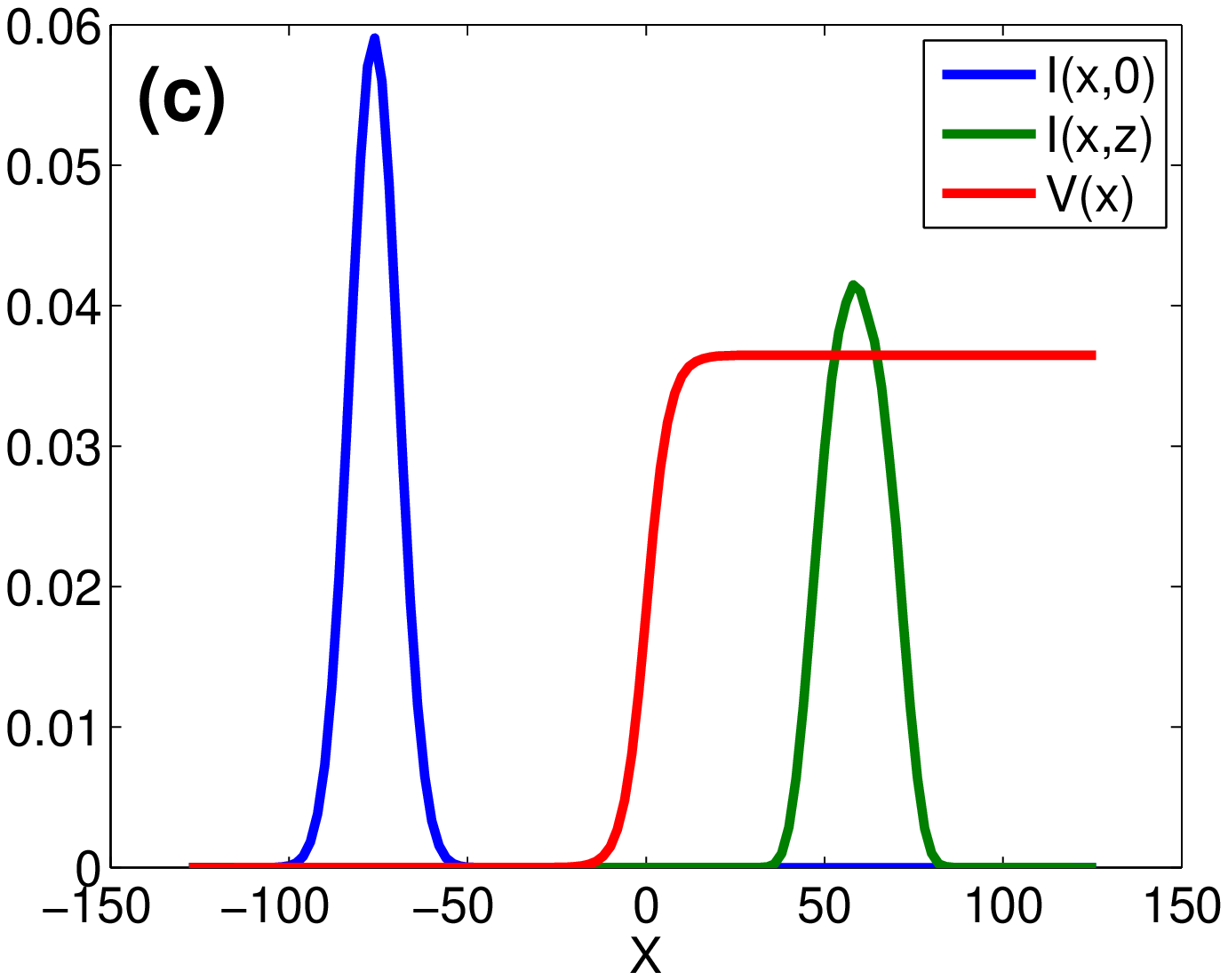}
    \includegraphics[width=0.23\textwidth]{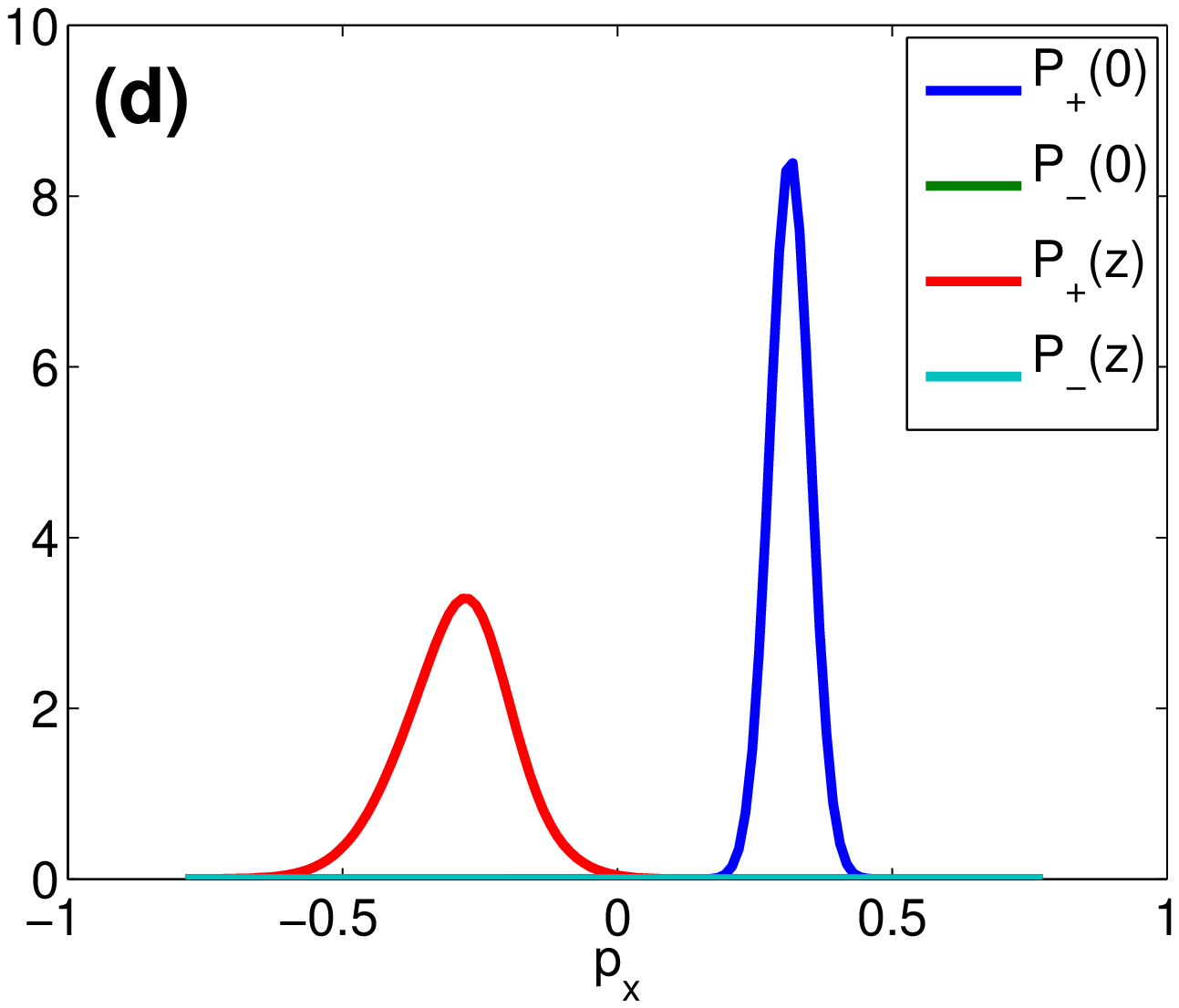}
    }%
    \caption {(a) The propagation constant and an illustration of the initial momentum distribution (dashed line) (b) The intensity of the wave packet in the $(x,z)$ plane. The dashed line represent the position of the potential step.
    (c) Illustration of the potential step (red) and the initial (blue) and final (green) intensities. 
    (d) The population of eigenstates of the initial (blue) and final wave packet.}
    \label{non linear klein U neq 0}
\end{figure}

Finally, we examine the effects of nonlinearity on one of the most remarkable phenomena associated with honeycomb lattices: Klein tunneling, where a wave packet tunnels into a potential step with probability 1 at normal incidence. It was found that in graphene, the coulomb interactions "strongly suppress Klein tunneling"  \cite{PhysRevB.80.235423}. Hence, it is interesting to study whether other types of interactions have such suppression effects or not. In this context, our type of nonlinearity represents not only EM waves in photonic lattices, but also interacting BEC's. 
We solve Eq.(\ref{1D dirac eq}) numerically in the presence of an additional smooth step-like potential. The initial wave packet is composed of modes associated with the second band, and it is initially located in the region of the higher refractive index. In spite of the presence of nonlinearity, we find that the wave packet is entirely transmitted, i.e., it tunnels in to the region of lower refractive index with probability 1, meaning that Klein tunneling is not suppressed at all. This is in contrast to quasi-particles in graphene where Klein tunneling is strongly suppressed  \cite{PhysRevB.80.235423}. Moreover, even when we relax the Dirac approximation, meaning taking into account  $\Ord(p^2)$ terms (Fig.  \ref{non linear klein U neq 0} (a)), the tunneling probability remains exactly 1 (Fig.  \ref{non linear klein U neq 0} (b) and Fig.  \ref{non linear klein U neq 0} (c) ). We verify this result by calculating the projection of the left and right moving plane waves  $\PP_{\pm} \equiv \lt| \bracket{p,\pm}{\psi(z)} \rt|^2 $. The calculation reveals that right and left movers do not exchange population (Fig.  \ref{non linear klein U neq 0} (d)). The fact that Klein tunneling at normal incidence is not effected by  $\Ord(p^2)$ terms was obtained already by  \cite{PhysRevLett.100.236801}, but that was for non-interacting waves only, whereas here we expand this finding to the nonlinear domain. However, since the "free" wave dynamics is strongly affected by the quadratic corrections, is it still surprising that Klein tunneling remains unaffected by it, and it is desirable to understand it. 

These findings can be explained in a fairly simple manner. The reflection amplitude is proportional to 
  $r\propto \bra{\beta_z,-p} W \ket{\beta_z,p}$, where  $W$ includes the external potential step and the nonlinear term. At normal incidence, the states  $ \ket{\beta_z,p}$ and  $\ket{\beta_z,-p}$ are  $\ket{-}$ and $\ket{+}$ respectively. Since the potential step is the same for both pseudo-spin components (there is no difference between the two sub-lattices), it is proportional to the identity operator. Moreover, since the incident wave packet is right (or left) moving, its two components are equal in magnitude, hence, the nonlinear term is proportional to the identity operator as well. Therefore, the reflection amplitude is proportional to the overlap of the states:  $r \propto \bracket{-}{+} $, that completely vanishes.

In conclusion, we studied quasi-1D waves' dynamics in honeycomb lattices in the presence of Kerr nonlinearity. We have put a special emphasis on Bloch waves at the vicinity of the Dirac points, and found that the quasi-1D wave packet dynamics is very different from the 2D waves' dynamics studied previously  \cite{PhysRevA.82.013830, nonlinear_dirac2}. In fact, we have found an infinite number of non-diffracting self-similar solutions that are insensitive to noise, and are therefore  immune to modulation instability. Moreover, we have shown that the most general wave packet is a superposition of left-moving and right-moving self-similar solutions, to a very good approximation. Finally, we reexamined Klein tunneling at normal incidence in the presence of nonlinearity, and found that, in contradistinction to other systems exhibiting Columb interactions, the tunneling probability is unaffected by Kerr-type nonlinearity.

\acknowledgments
We thank Yaakov Lumer for useful discussions and revising the manuscript.
This work is supported by an advanced grant from the ERC.

\end{document}